# Types and Forms of Emergence

Jochen Fromm

Distributed Systems Group,
Electrical Engineering & Computer Science,
Universität Kassel, Germany

**Abstract.** The knowledge of the different types of emergence is essential if we want to understand and master complex systems in science and engineering, respectively. This paper specifies a universal taxonomy and comprehensive classification of the major types and forms of emergence in Multi-Agent Systems, from simple types of intentional and predictable emergence in machines to more complex forms of weak, multiple and strong emergence.

## 1. Introduction

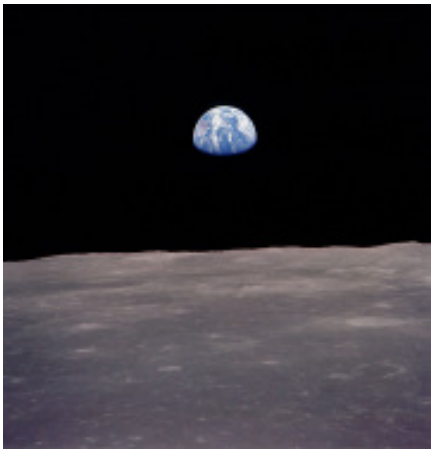

The emergence of order and organization in systems composed of many autonomous entities or agents is a very fundamental process. The process of emergence deals with the fundamental question: "how does an entity come into existence?" In a process of emergence we observe something (for instance the appearance of order or organization) and ask how this is possible, since we assume causality: every effect should have a cause. The surprising aspect in a process of emergence is the observation of an effect without an apparent cause. Although the process of emergence might look mysterious, there is nothing mystical, magical or unscientific about it.

If we consider the world of emergent properties, the deepest mysteries are as close as the nearest seedling, ice cube, grain of salt or pile of sand, as Laughlin explains in his book [Laughlin05]. It is doubtful that the ultimate laws can be found at inconceivable high energies or extreme scales, if we do not understand things at our own scale well enough. In other words we must step back and look at the patterns and the interactions of everyday objects to discover the nature of our universe.

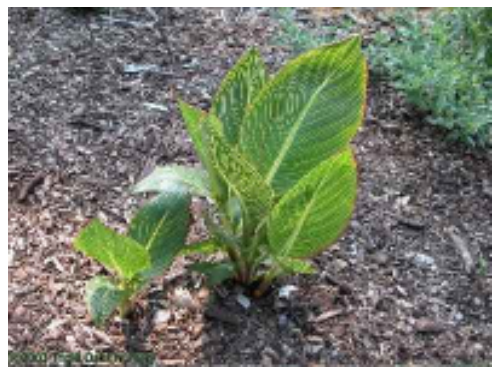

Emergent properties are amazing and paradox: they are very fundamental and yet familiar. Emergent phenomena in generated systems are according to John H. Holland typically persistent patterns with changing components [Holland98], i.e. they are changeless *and* changing, constant *and* fluctuating, persistent *and* shifting, inevitable *and* unpredictable. Moreover an emergent property is a part of the system and at the same time it is not a part of the system, it depends on a system because it appears in it and is yet independent from it to a certain degree. According to the Stanford Encyclopedia of Philosophy, "emergent entities (properties or substances) 'arise' out of more fundamental entities and yet are 'novel' or

'irreducible' with respect to them"[1]. Because true emergent properties are irreducible, they can not be destroyed or decomposed – they appear or disappear instead. In this sense they may seem to be indestructible and are potentially the only things that really exist, but if they are examined too closely - if we take a deeper look at the components of the system - they do not exist at all and often vanish into nothing. The paradoxes arises mainly because we are often only able to see a part of a complex system, if we consider only the microscopic *or* the macroscopic level, but not both at once, or if we see only the system *or* the environment, but not both. It is difficult to comprehend a complex system with multiple levels and scales.

> Emergence is paradox: emergent properties are often changeless *and* changing, constant *and* fluctuating, persistent *and* shifting, inevitable *and* unpredictable, dependent *and* independent from the system in which they arise.

Unforeseeable failures and unexpected faults in software and hardware systems are a special, undesirable form of emergence. As Duncan J. Watts says "the best maintenance procedures in the world can't guarantee to prevent faults that aren't yet known to exist [...] failure will happen despite our best efforts" [Watts03]. It is necessary to understand the process of emergence in complex systems in order to create new forms of complex and robust systems, which are prepared for the occurrence of errors and prevent failure as much as possible. The knowledge of the different types of emergence is essential if we want to understand and master complex systems in science and engineering, respectively.

Self-organization among social animals usually involves an emergence 'trick': the **pheromone trick** (mark items of interest with a pheromone field) and the **flocking trick** (stay near the group but not too close to your neighbours). These two simple tricks lead to colonies of ants, piles of termites, swarms of bees, flocks of birds, herds of mammals, shoals/schools of fish, and packs of wolves. Are there other tricks or is this the end of it?

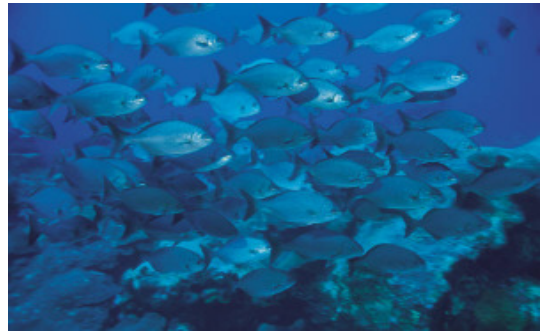

A clear formal description would be helpful to answer this question. Aleš Kubík [Kubik03] and Yaneer Bar-Yam [BarYam04] tried to do this with grammars, formal languages and mathematics. Yet the term "emergence" is too general to cover it with a jargon of a certain science: "it is unlikely that a topic as complicated as emergence will submit meekly to a concise definition" [Holland98]. Emergence is hard to capture with a model or a theory, just because during an emergence process new and unpredictable entities appear, which are governed by their own laws. It is associated with the necessity to define new categories, concepts and descriptive terms. The problem is as Robert I. Damper argues "How will we avoid doing this afresh for each and every case?" [Damper00].

Since formal descriptions with grammars, formal languages or mathematics are apparently inappropriate and unsuitable to solve this difficult problem, at least a complete taxonomy or systematic classification would be helpful. This paper is an attempt to propose a standard classification, specification and taxonomy for the different types and forms of emergence.

---

[1] http://plato.stanford.edu/entries/properties-emergent/

## 2. Definitions and Attributes

### 2.1 Definitions

Since emergence is an ambiguous word, it is important to start with a clear definition. In the following, emergent and emergence are defined like this: a property of a system is **emergent,** if it is not a property of any fundamental element, and **emergence** is the appearance of emergent properties and structures on a higher level of organization or complexity (if "more is different" [Anderson72]). This is the common definition that can be found in many introductory text books on complex systems [Flake00, BarYam97].

The Oxford Companion to Philosophy [Honderich95] defines **emergent properties** as unpredictable and irreducible: "a property of a complex system is said to be 'emergent' just in case, although it arises out of the properties and relations characterizing its simpler constituents, it is neither predictable from, nor reducible to, these lower-level characteristics".

The Cambridge Dictionary of Philosophy [Audi95] distinguishes between structures and laws, between descriptive and explanatory emergence. **Descriptive emergence** means "there are properties of 'wholes' (or more complex situations) that cannot be defined through the properties of the 'parts' (or simpler situations)". **Explanatory emergence** means "the laws of the more complex situations in the system are not deducible by way of any composition laws or laws of coexistence from the laws of the simpler or simplest situations".

The Cambridge Online Dictionaries[2] say *to emerge* means "to appear by coming out of something or out from behind something" or "to come to the end of a difficult period or experience", whereas emergence is simply "the process of appearing". The words emergence and to emerge have the Latin origin *emergere* which means to rise up out of the water, to appear and to arrive. The Latin verb *emergere* comes from *e(x)* "out" + *mergere* "to dip, plunge into liquid, immerse, sink, overwhelm", i.e. **to emerge** is something like the opposite of **to merge**. Emergence is derived from the present participle of emergere, *emergens*. Whereas merging means the combination, immersion, fusion of two separate things, *e*merging means the opposite.

### 2.2 Attributes and Criteria

The concepts of **explanation**, **reduction**, **prediction** and **causation** are central to a deeper understanding of emergence [Kim99]. Important characteristics of emergent properties are "unexplainability", "irreducibility", unpredictability and feedback, although the degrees in different systems vary. Francis Heylighen has proposed the following classification criteria for emergence [Heylighen91]

- amount of variety in the created system
  (i.e. in the possible states of the emergent system)
- amount of external influence (during the process of emergence)
- type of constraint maintaining the system's identity (absolute or contingent)
- number of levels, multi-level emergence (one level, two levels, or multiple levels)

This attributes are unfortunately not completely independent: a multi-level emergence process has certainly a much higher variety than an emergence process on a single level.

---

[2] http://dictionary.cambridge.org/

# 3. Classifications and Taxonomies

## 3.1 Philosophical Taxonomies

David J. Chalmers distinguishes between weak and strong emergence [Chalmers02]. His **strong** emergence is not deducible even in principle from the laws of the low-level domain, while **weak** emergence is only unexpected given the properties and principles of the low-level domain. Mark A. Bedau distinguishes between three kinds of emergence: nominal, weak and strong [Bedau02]. He uses weak and strong in the same sense as Chalmers, and adds the notion of **nominal** emergence, which corresponds to the general definition of an emergent property above: nominal emergence is the appearance of a macro property in a system that cannot be a micro property. William Seager [SeagerDrafts] emphasizes two kind of emergence: **benign** and **radical**. He distinguishes between supposedly 'benign' or acceptable emergence - if we can find a descriptive or explanatory scheme which provides a useful kind of shorthand notation for describing the behavior of a system, for example the pressure and temperature of a gas - from a 'radical' emergence. His radical emergence corresponds roughly to the strong emergence of Bedau and Chalmers.

## 3.2 Taxonomies for Particles

Yaneer Bar-Yam [BarYam04] distinguishes between four types of emergence. He gives the following short description, which is based on "particles" and "ensembles" rather than agents and groups:

- Type A Emergent Behavior (Micro to macro)
    Type 0 (Parts in isolation without positions to whole)
    Type 1 (Parts with positions to whole - weak emergence)
    Type 2 (Ensemble with collective constraint - strong emergence)
    Type 3 (System to environment relational property - strong emergence)
- Type B Dynamic emergence of new types of systems "new emergent forms"

Type 0 is simply the "parts isolated vs. parts joined" emergence, similar to benign and nominal forms. Type 1 and Type 2 correspond roughly to the weak and strong emergence, although his notion of strong emergence is a bit idiosyncratic ("A strong emergent property is a property of the system that cannot be found in the properties of the system's parts or in the interactions between the parts"). For strong emergence, he emphasizes correctly the existence of intermediate components or ensembles between the global level of the system and the local level of the agents. Type 3 classifies the emergent behavior of systems which arise out of the interaction with the environment.

## 3.3 Taxonomies for CA

There are two general classifications or taxonomies for CA, Wolfram's and Eppstein's classification. Stephen Wolfram's four CA classes are [Wolfram84] homogeneous (Class I ) regular, periodic (Class II), chaotic (Class III) and complex (Class IV). David Eppstein's simpler classification[3] was formulated to overcome difficulties with Wolfram's scheme. He proposes three major classes:

- Contraction impossible (negative feedback)
- Expansion impossible (positive feedback)
- Both expansion and contraction possible (neg. and pos. feedback)

---

[3] http://www.ics.uci.edu/~eppstein/ca/wolfram.html

# 4. Taxonomy

## 4.1 A New Taxonomy

The taxonomy in this paper is based on different feedback types and the overall structure of causality or cause-and-effect relationships, which fits perfectly to the classification of Eppstein for CA. Causal relations and cause-effect relationships are a natural order for all forms and types of emergence, because the cause is normally the unclear point in emergence: *emergence is an effect or event where the cause is not immediately visible or apparent*. A question about a process of emergence is always a question of causation and causality – the search of a hidden cause for an apparent effect. The different types of emergence can be classified roughly through four types or classes.

**Type I Simple/Nominal Emergence
without top-down feedback**
Type Ia Simple Intentional Emergence
Type Ib Simple Unintentional Emergence

**Type II Weak Emergence
including top-down feedback**
Type IIa Weak Emergence (Stable)
Type IIb Weak Emergence (Instable)

**Type III Multiple Emergence
with many feedbacks**
Type IIIa. Stripes, Spots, Bubbling
Type IIIb. Tunneling, Adaptive Emergence

**Type IV Strong Emergence**

**Type I** contains no feedback at all, only "feedforward" relationships[4]. The major characteristic of **Type II** is simple feedback: (a) positive or (b) negative. Multiple feedbacks, learning and adaptation are important for **Type III**. John H. Holland said about emergence in adaptive systems: "Any serious study of emergence must confront learning" [Holland98].

**Type III** appears in very complex systems with many feedback loops or complex adaptive systems with intelligent agents. It is the class with a large amount of external influence during the process of emergence (the internality/ externality dimension of Heylighen and Bar-Yam's system to environment relational property). **Type IV** emergence is characterized in the words of Heylighen [Heylighen91] by *multi-level* emergence and a huge amount of variety in the created system, i.e. the number of possible states of the emergent system is astronomical due to combinatorial explosion. It is the form of emergence which is responsible for structures on a higher level of complexity which cannot be reduced, even in principle, to the direct effect of the properties and laws of the elementary components.

Stephen Jones has used a similar approach to define a comprehensive taxonomy [Jones02]. He uses feedback relations to propose a taxonomy of emergence, and distinguishes between first order (only "feedforward" relations), second order (feedback relations), and third order ("mutualistic feedback" relations) forms of emergence. Unfortunately, the third category is unclear and seems to contain all other cases. Moreover strong emergence seems to be completely missing in his taxonomy. He further tries to apply the different feedback forms known from cybernetics to differentiate his second category of emergence: his taxonomy distinguishes between feedback with error values (in which the output is filtered and limited before it is returned to the input) and feedback without error values (in which the pure output is returned to the input). This is not a bad idea, but the explanatory power in the case of emergence is doubtful.

---

[4] The idea to a type I class is due to Prof. Abbott, see for example [Abbott05a,Abbott05b]

## 4.2 Different perspectives

The classification can also be seen from a different perspective, if it is specified in terms of constrained generating processes or roles: **type I** corresponds to fixed roles, **type II** to flexible roles, **type III** to the appearance of new roles and the disappearance of old ones, **type IV** to the opening of a whole new world of new roles. Another classification possibility is to use different levels of prediction: intentional emergence of **type I** is predictable, weak emergence of **type II** is predictable in principle (though not in every detail), multiple emergence of **type III** is chaotic or not predictable at all, strong emergence of **type IV** is not predictable in principle.

| Type | Name | Roles | Frequency | Predictability | System |
|---|---|---|---|---|---|
| I | **Nominal or Intentional** | fixed | abundant | predictable | closed, with passive entities |
| II | **Weak** | flexible | frequent | predictable in principle | open, with active entities |
| III | **Multiple** | fluctuating | common - unusual | not predictable (or chaotic) | open, with multiple levels |
| IV | **Strong** | new world of roles | rare | not predictable in principle | new or many systems |

Common elements of all types are boundaries, feedbacks and jumps/leaps. As the name emergence suggests, the emergence of something is always possible at a clear **boundary** of a system, and usually a **jump** or **leap** to a new level occurs. Moreover nearly all interesting types of emergence involve some form of **feedback** and one or more feedback loops: scale-preserving (peer-to-peer) feedback in Type Ib, scale-crossing (top-down) feedback in Type II, and multiple feedbacks in the other types.

| Type | | Boundaries | Feedbacks | Jump/Leap |
|---|---|---|---|---|
| I | Ia | agent-system boundary (only in one direction) | no feedback but absolute commands or constraints | intended or static jump to higher level of organization |
| | Ib | agent-agent boundary | scale-preserving (peer-to-peer) feedback | fluctuations, no jump to significant higher level of organization |
| II | II | agent-group or agent system boundary (in both directions) | scale-crossing (top-down) feedback, positive *or* negative | dynamic jump to higher level of organization |
| III | IIIa | agent-group or agent system boundary (in both directions) | scale-crossing (top-down) feedback, positive *and* negative | dynamic jump to higher level of organization |
| | IIIb | large fitness barriers in complex evolutionary systems | multiple feedbacks in a system | quantum leap in complex adaptive system |
| IV | IV | boundary between different evolutionary systems, *barrier of relevance* | all of the above, incl. feedback between different systems | gateway or quantum leap in evolution to new (evolutionary) system |

# 5. Examples

## 5.1 Type I. Simple Emergence without top-down feedback

**5.1 a) Simple Intentional/Nominal Emergence.**

The only constrained generating role or top-down process in **type Ia** is the intentional design of a machine: a specific and fixed role is assigned to each part, and this role does not change in the course of time. The behavior of each part is always the same, it is independent of the other parts' states, the global state of the system and the environment. The function of a complicated or designed machine with **type Ia** emergence is an intended emergent property of the controlled and planned interaction of the individual parts and components:

- The function of a machine is an emergent property of its components
- The function of a software system is an emergent property of the underlying code
- The semantic information of a sentence is an emergent property of the sounds and words in the sentence that depends on how they are arranged

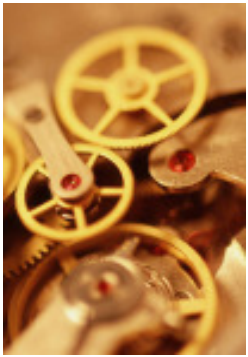

It is typical for emergence in complex systems that new roles are assigned to the agents and actors. In this type, no other form of role assignment is allowed. An example for a typical role is "explorer", "follower" or "transporter" during foraging in ant colonies. In a computer program or in a complicated system, every component and object has only a certain well-defined role and is governed by well understood rules. Certainly the gears of a watch have a role in the watch that they did not have before, but once they are built in, these roles do not change anymore.

If the behavior of each part is *independent* of the other parts' states, then the behavior of a machine is deterministic and predictable: such a system is according to Ashby [Ashby62] 'self-organizing' in the sense that it changes from 'parts separated' to 'parts joined'. **Type Ia** emergence is typical for ordinary machines as clocks or steam engines. Ashby defined a machine as follows: "A 'machine' is that which behaves in a machine-like way, namely, that its internal state, and the state of its surroundings, defines uniquely the next state it will go to." [Ashby62].

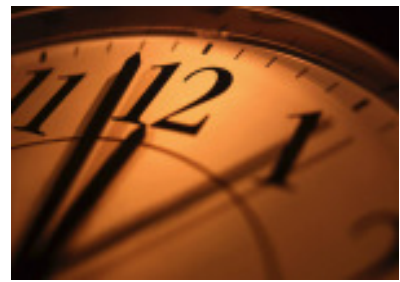

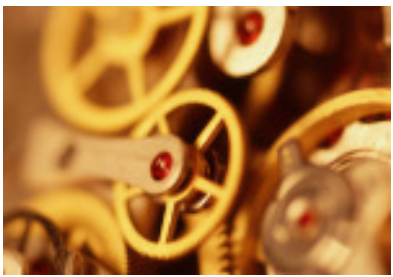

This form of nominal or intentional emergence is not new. It has been used for centuries. If we create such a system, we have a certain intention, aim or plan. We debug and change the system until the function that emerges from the assembly of parts and components matches exactly that single purpose. Yet intended emergence has a drawback, as we can observe in large systems with complete top-down control (for example a planned command economy).

A typical drawback of a system with **type Ia** emergence is brittleness and lack of flexibility or adaptability. Well-defined and planned objects are 100% reliable and act always dutiful to fulfil their obligations. If there are no redundant components, a defect in a single component can bring the entire system to a halt.

### 5.1 b) Simple Unintentional Emergence

The statistical quantities and properties of a number of identical particles are an example of unspectacular **Type Ib** emergence (if they depend on a relation of the particles to each other):

- thermodynamic properties like pressure, volume, temperature
- characteristic path lengths and clustering coefficients in networks
- emergent physical properties like an avalanche, a wave-front or a slope of a sand-pile

**Type Ib emergence** appears in a system with many loosely coupled, disorganized and equal elements, which possesses certain average properties as temperature or pressure. It can barely be called emergence, because it describes the properties of a large number of agents as the sum of an average property. An average quantity alone is not an emergent feature. Yet statistical quantities which define properties of an aggregation can be regarded as simple emergent properties, *if they depend on a relation of the particles to each other*, i.e. if they do not make sense for a single particle. A quantity like pressure can not be applied to a single particle, a single grain of sand has no slope, and a single node in a network has no path length. Most averages are averages of properties that apply to the pieces (e.g. average salary of PhDs in Computer Science) are not an example of **type Ib emergence**, because they violate the notion that a property is emergent only if it applies to the whole and not to the parts.

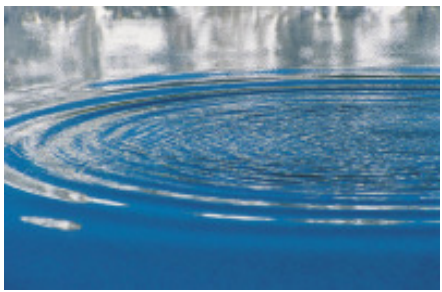

Even if there is no top-down feedback, there can be a simple feedback on the scale of the agents or particles similar to a chain reaction or a domino effect: a particle is excited, which influences a new particle, then a new particle is excited etc. This simple form of scale-preserving (peer-to-peer) feedback leads to waves, chain-reactions, cascades, avalanches, etc.

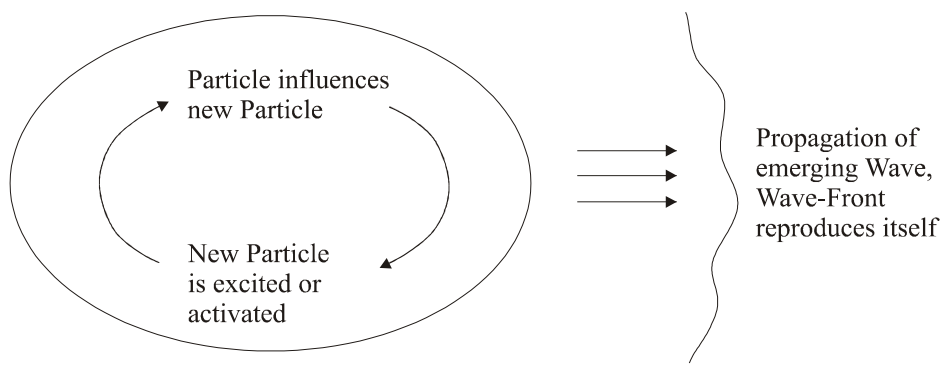

## 5.2 Type II. Weak Emergence including top-down feedback

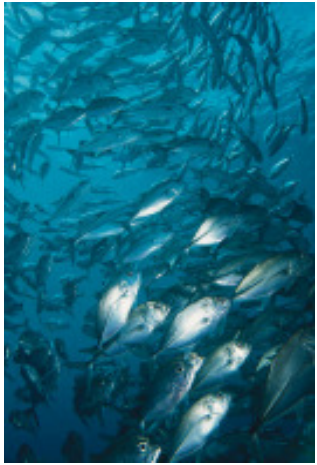

**Type II** emergence includes a top-down feedback from the macroscopic to the microscopic level. If we consider a Multi Agent System (MAS), we can identify various levels of spatial resolution (agent-group, entity-system, individual-collective, particle-swarm…). On the low microscopic level of the agent, many individual entities or agents interact locally with each other. This interaction results in a new, usually unpredicted pattern which appears at a higher level. On the high macroscopic level of the group or whole system, we notice unpredicted patterns, structures or properties - emergent phenomena - which are not directly specified in the interaction laws, and which in turn influence the low-level interactions of the entities via a feedback process.

A swarm or shoal of fish for instance is an emergent property which influences the motions of each participating animal. It can be explained and described by **type II** emergence, which is also known as weak emergence. M.A. Bedau has defined weak emergence as follows: a macroscopic state or property is weakly emergent if it can be derived from the microscopic dynamic but only by simulation [Bedau97]. It is predictable in principle, but not in every detail. The difficulties in prediction and derivation of weakly emergent properties arises of course from the top-down feedback process, which is among philosophers known as 'downward causation' [Bedau02]. In weak emergence, there is no unique direction of causality from the microscopic to the macroscopic level as in simple forms of nominal or intentional emergence. There are causal relations in both directions.

There are two basic forms of interaction which lead to weak emergence (see [Küppers96]):

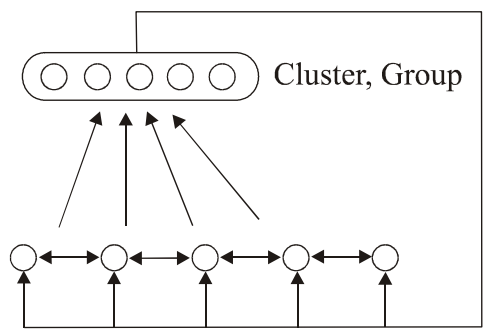
Direct Interaction

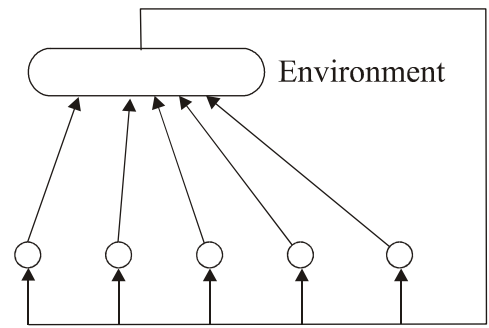
Indirect Interaction through environment

**Direct Interaction.** The agents or entities of the system interact directly with each other. The interactions lead to the formation of groups and clusters, which influence in turn the behavior of the agents or entities. The feedback is only possible if the agents can distinguish between different scales, for instance microscopic objects have a repulsive effect whereas macroscopic objects have an attractive effect. This happens in the **flocking trick** mentioned earlier (stay near the group but not too close to your neighbours). Examples are swarms of bees, flocks of birds, herds of mammals, shoals/schools of fish, and packs of wolves.

**Indirect Interaction.** The agents or entities of the system change the state of the total system and the environment through their individual behavior. The changes in the environment influence in turn the behavior of the agents or entities. The indirect feedback is possible if the agents can manipulate the environment through long-lasting local changes with global and persistent effects, as in the **pheromone trick** mentioned earlier (mark items of interest with a pheromone field), or if they act nearly synchronously so that the effect of the individual actions is multiplied. Examples in natural systems are colonies of ants and piles of termites, in social systems migration processes.

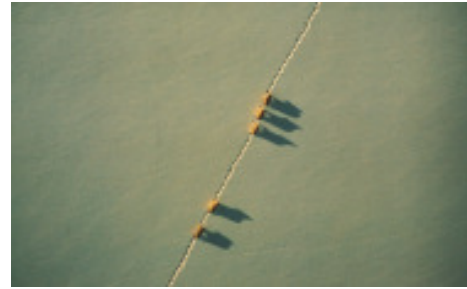

There are of course also mixtures of both forms. The groups and clusters which arise through direct interaction in the first case can influence the state of the environment. In the second case, indirect feedback through the environment often leads to reinforcement of group formation, esp. in social rituals with synchronous actions which melt a group together.

Additional to direct and indirect forms of interactions and feedbacks, there are two different types of weak emergence according to the two different types of possible feedback: negative or damping feedback leads to stable forms of weak emergence, positive or amplifying feedback results in instable forms as short-lived fads, bubbles and buzz.

### 5.2 a) Weak Emergence (Stable)

Usually the feedback is negative and imposes a constraint on the actions of the agents. Weak unintentional emergence of **Type IIa** is the classic form of emergence which includes bottom-up influences and top-down feedback from the group or the environment. It is related to swarm intelligence, group formation and stigmergy, and to the flocking and the pheromone trick mentioned earlier. Examples are

- foraging behavior of ants colonies
- flocking behavior of fish and birds
- a liquid or fluid is an emergent property of molecules
- forms of self-organization in the Internet, for example in the World Wide Web (WWW) and WIKIPEDIA, in Open-Source projects as Linux and Mozilla
- optimal prices of goods in an economy and free markets according to the law of supply and demand

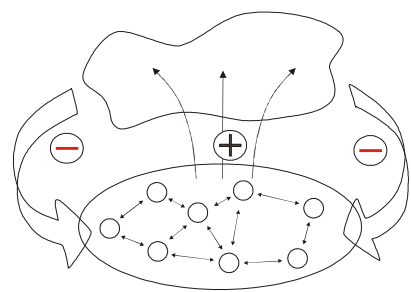

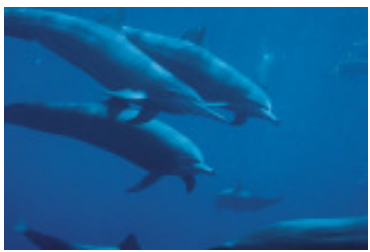

In stable forms of weak emergence there is a balance between exploration, diversity and randomness (through bottom-up influences) on the one hand and exploitation, unity and order (through top-down constraints) on the other hand. **Type IIa** emergence is based on two levels which are connected by a bottom-up and a complementary top-down process.

Diversity and 'exploration' arises out of the 'creative' bottom-up process due to the autonomy of the components and their unique contexts. Unity and 'exploitation' arises out of the 'constraining' top-down feedback processes, which imposes a constraint on the components or agents — similar to "peer pressure" and forces them to adjust their behavior or to occupy a certain role. The complex forms of organization that we observe in a system with emergent properties are due to a balance of diversity and unity, of exploration and exploitation, of 'creative' bottom-up process and 'constraining' top-down feedback. Ants in a swarm explore every direction due to their different individual contexts and random influences. Yet they have a collective goal and are forced to follow their own pheromone trails. Birds in a flock fly in different directions to avoid collisions, but at the same time stay close together, try to match the neighbors' velocity and steer to the perceived center of the flock.

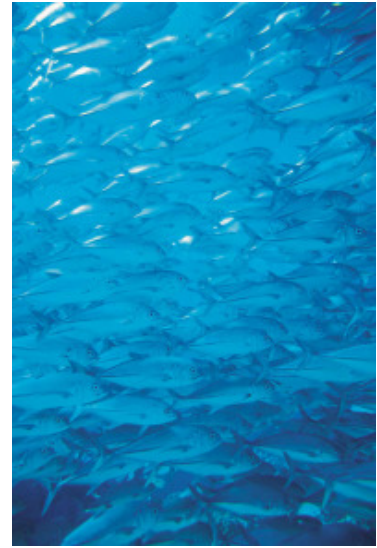

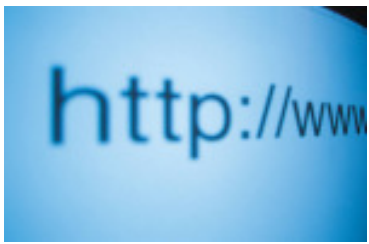

The World Wide Web (WWW) mirrors the huge diversity of the world population. As Flake et. al say, "Millions of individuals operating independently and having a variety of backgrounds, knowledge, goals, and cultures author the information on the Web" [Flake02]. This diversity is balanced by the unifying standards and constraints of the W3C and other consortiums (HTML, HTTP,etc.).

This form of self-organization can be very powerful, as the continuous success of Linux, Mozilla and the Firefox Browser, WIKIPEDIA and countless other open and free collaborative projects around the world show. WIKIPEDIA relies on the diversity of its participants and contributors. It would not work if it should be created with a clone army where every participant is exactly identical to its peer. Yet there is a unifying force with preserves unity: every participant uses the same simple editor and obeys the same easy rules.

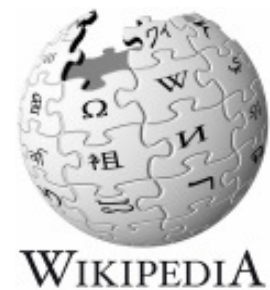

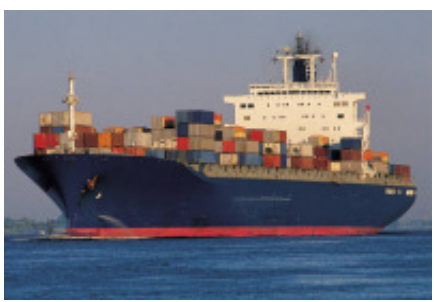

Optimal pricing of goods in an economy and free markets emerges from the interaction of agents obeying the local rules of commerce and *the law of supply and demand*. It is based on negative, stabilizing feedback: high demand or undersupply of resources cause high prices, high prices in turn reduce the demand and increase the resources. Low demand or oversupply of resources cause low prices, low prices in turn increase the demand and reduce the resources. Thus the demand is changed through negative feedback until a partial equilibrium is reached. The law of supply and demand assumes that the market is perfectly competitive (no actor has enough power to influence the prices directly) and that the agents act rationally, which is not always the case. Someone who only imitates others does not act rationally, i.e. his actions are not optimal. He will for example pay a price which is not adequate or appropriate for a product.

## 5.2 b) Weak Emergence (Instable)

Feedback can also be positive. Drug addiction or economic inflation are (negative) examples for positive feedback: high wages lead to high product prices, high prices increase the cost of living, and high costs of living increase wages. In drug addiction, high doses activate the brain's reward systems and result in repeated consumption, repeated consumption leads to habituation which modifies the reward system and leads to even higher doses.

It is well-known that imitation is one of the primary factors influencing the creation of fads, fashion cycles, and abrupt collective opinion shifts [MichardBouchaud05, Tassier04]. As Duncan Watts has observed [Watts03], our decisions are frequently based on the actions of others. Before we have to decide something important (buy a share on a financial market, etc.), we often look what other people are doing, we imitate others. Yet it is possible that others do the same, and do not really know what to do, either. When everyone is trying to make decisions based on the actions of others, collectives may fail to aggregate information, and positive feedback loops and mutual reinforcement can appear. For instance short-lived fads: if an item is frequently bought, it is regarded as valuable, and if is valuable, many consumer will buy it, if they only look what others are doing. If distributed decision making fails like in this case, then small fluctuations from equilibrium can lead to giant and dramatic effects: buzz, bubbles and short-lived fads, or cascades and fast growing avalanches:

- **Bubbles** and **crashes** in stock markets and financial markets
- **Fads**, crazes and skewed distributions in cultural markets
- Sudden **explosions** of social unrest (e.g. East Germany, Indonesia, Serbia)
- Changes in previously stable social norms
- **Bandwagon effect**
- **Celebrity effect** (someone who is famous principally for being well-known)
- **Buzz** in the news
- cluster formation in economic and social migrations (ghettos, slums)
- Path-dependent evolution of lock-in states: the VCR / Keyboard / PC / OS market

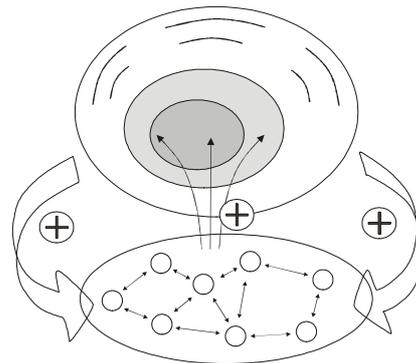

A trend follower only imitates others instead of really estimating the value of things. He loses the contact to reality, just as the bubble that eventually follows, if all others do the same. **Type IIb** emergence is a form of undesirable, negative emergence through positive feedback, until the bubble explodes or vanishes due to exponential growth. The price of an explosive phase is the following burnout phase: a financial bubble explodes and ends in a crash, a cultural fad ends in oversaturation and a political revolution ends often in political indifference.

Certain forms of inappropriate buzz in the news are similar to the emergence of bubbles in stock markets or short-lived fads. As Steven Johnson reports in his book "Emergence" (see chapter 4 in [Johnson02]), media has a tendency for self-amplification. The discussion if you should cover certain things in the media (for example in reports about O.J. Simpson, Bill Clinton or Princess Di's death, if the media is involved itself in the case) is a self-coverage and leads to positive feedback: self-coverage leads to situations where the coverage of a story begets more coverage, because a story is regarded as important if it is frequently covered.

The decisions of TV stations what to broadcast are frequently based on the actions of others, they often look what other stations are doing. When everyone is trying to make decisions based on the actions of others, collectives may fail to aggregate information, and distributed decision making does not work. The possibility of a cascade is certainly determined by the underlying communication network. Complex scale-free or small-world networks are especially vulnerable to cascades, because on the one hand every node is separated from every other node only by a few connections, thus information cascades can spread fast. On the other hand such a network contains many easily influenceable individuals, because the average degree of connectivity is low, much lower than the amount of nodes in the whole network. Individuals which are easily influenceable are not the most connected individuals or opinion makers. Thus scale-free or small-world networks are especially vulnerable to cascades due to collective decision making.

Cluster formation in economic and social migrations (high-tech regions, slums, ghettos) is another example of positive feedback. Good locations attract interesting firms, and interesting firms make the location better. This is one factor that leads to the emergence of economic clusters and high-tech regions like Bangalore or the Silicon Valley. Poor living conditions in a district (increased violence, use of weapons and drugs) lead to migration of the middle class, enhanced migration of the middle class can worsen to conditions. This results in the formation of slums and ghettos.

This sort of positive feedback process is known in the economy also as "increasing returns". The concept of positive feedbacks and increasing returns in the economy was illustrated well by W. Brian Arthur [Arthur90, Arthur94]. VHS tapes, IBM PCs, QWERTY-Keyboards, DOS and Microsoft Windows captured the entire market completely through the same path-dependent positive feedback process. The process is path-dependent, because small fluctuations or accidents at the beginning can be reinforced through the positive feedback process to large, unpredictable deviations, until they are finally conserved for a long time in the resulting frozen lock-in state, where one product dominates and allocates the entire market (German Keyboards replace for example Y with Z, they read QWERTZ instead of QWERTY). The reason for positive feedback and self-reinforcing mechanisms is again imitation: "I need a VHS / IBM PC / DOS / Windows product because everyone else buys / has / uses it".

Positive feedback is probable in markets where imitation is useful, where compatibility and data exchange is very important, and where the costs for obtaining, installing and learning to use a new technology are high. As W. Brian Arthur noticed [Arthur90] "The parts of the economy that are resource-based (agriculture, bulk-goods production, mining) are still for the most part subject to diminishing returns (negative, stabilizing feedback)" whereas "The parts of the economy that are knowledge-based, on the other hand, are largely subject to increasing returns (positive feedback).".

## 5.3 Type III. Multiple Emergence with many feedbacks

**5.3 a) Stripes, Spots, Bubbling - Emergence with multiple feedback**

Real financial or stock markets, and many pattern formation processes in nature show neither pure Type IIa nor Type IIb behavior. They are often based on a combination of both. The idea that short-range activation (positive-feedback) coupled to long-range inhibition (negative feedback) can cause pattern formation is not new. In the context of chemical reactions, Alan Turing published a similar idea already 1952 [Turing52]. Today systems with **short-range positive-feedback** and **long-range negative feedback** are known as activator-inhibitor systems and belong to reaction-diffusion systems.

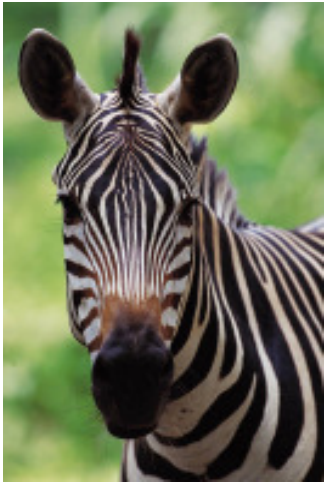

In reaction-diffusion systems, an activator "auto"-catalyzes its own production, but also activates an inhibitor, which diffuses more rapidly. For short-ranges the positive feedback prevails, for long-ranges the negative feedback takes hold. These systems are responsible for many types of biological pattern formation [Ball99], especially **stripes** and **spots** in animals coat patterns (leopard spots, jaguar pigments, zebra stripes,…) through activating and inhibiting melanocytes, cells that produce colored pigment [Murray02]. The process can also be simulated in simple Cellular Automata, for example through the rule "add up the number of immediate neighbors who are white minus 1/10 of the number of white neighbors in 20x20 blocks to the left and right of its immediate neighbors. If the sum is greater than one the cells becomes white, otherwise it becomes black"[5].

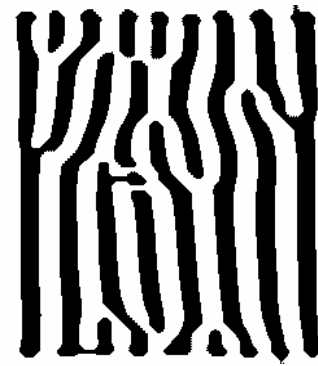

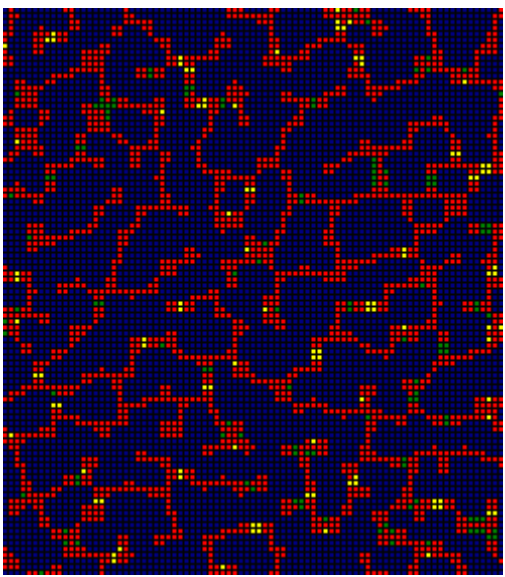

Patterns with spots and rings like the ones on jaguars and leopards can be produced by similar rules. Another example for short-range positive-feedback and long-range negative feedback is the iterated Prisoners' Dilemma on the lattice [NowakMay92]. Their evolutionary game consists only two kinds of players: those who always cooperate, and those who always defect. Every lattice site is occupied by a player. In each round, the players play the Prisoners' Dilemma game with their immediate neighbors, and the score of the player is the sum of the pay-offs in these encounters. At the end of the round (or generation), each lattice site is occupied by the player with the highest score among the previous owner and the neighbors. The result is that networks of "defectors" and cheaters (red) emerge in a sea of cooperators (blue). Cooperators support and reinforce each other, but if the clusters of cooperators grow too large, the conditions for defectors are getting better: red cheaters who exploit the cooperators spread and limit the size of blue clusters.

---

[5]see http://grace.evergreen.edu/artofcomp/examples/zebra/Zebra.html   A similar example for a discrete version or idealization of a reaction-diffusion process is also given by Stephen Wolfram in his NKS book on page 427 [Wolfram02]

Real financial or stock markets show often chaotic und unpredictable behavior, produce fractal and multifractal structures, and there are unpredictable runs of successive losses and crashes, but also repeated speculative bubbles of all sizes, see for example [Sornette03]. A system like the stock market which has both positive and negative feedback mechanisms can show oscillating and chaotic behavior.

You can find frequently a combination of **short-term positive feedback** (blind imitation) and **long-term negative feedback** (careful consideration). When stocks are rising, the belief that further rises are probable gives investors, imitators and trend-followers an incentive to buy, which will rise the price even more (positive feedback); but eventually the increased price of the shares, and the knowledge that there must be finally a peak after which the market will fall and prices will drop, ends up deterring buyers (negative feedback). If the stocks are falling the process is similar. Once the market begins to fall regularly, some investors may expect further declines and refrain from buying (positive feedback), but others may buy because stocks become more and more of a bargain (negative feedback). The combination of short-term positive feedback and long-term negative feedback leads to "bubbling", fluctuating or turbulent sequences of bubbles. Short-lived bubbles of all sizes appear that rise and expand as quickly as they disappear again. A bubble is triggered by a fluctuation or irregularity, It expands through positive feedback, and shrinks again through negative feedback.A good model for this bubbling behavior is the cellular automata model "Larger than Life", an extension of Conway's Game of Life to a larger radius. The definition of Conway's Game of Life is simple: A birth occurs at x if the population within its neighborhood equals 3. Site x stays occupied/alive if the count is in [2..3]. The "Bugs" rule for "Larger than Life" is very similar: A birth occurs at x if the population within its neighborhood (x included or not) lies in the interval [34..45] (positive feedback). Site x stays occupied if the count is in [34..58] (negative feedback for very small and very large values, due to loneliness and overpopulation).The screenshots made with the CA simulator MJCell from Mirek Wójtowicz show a series of bubbles.

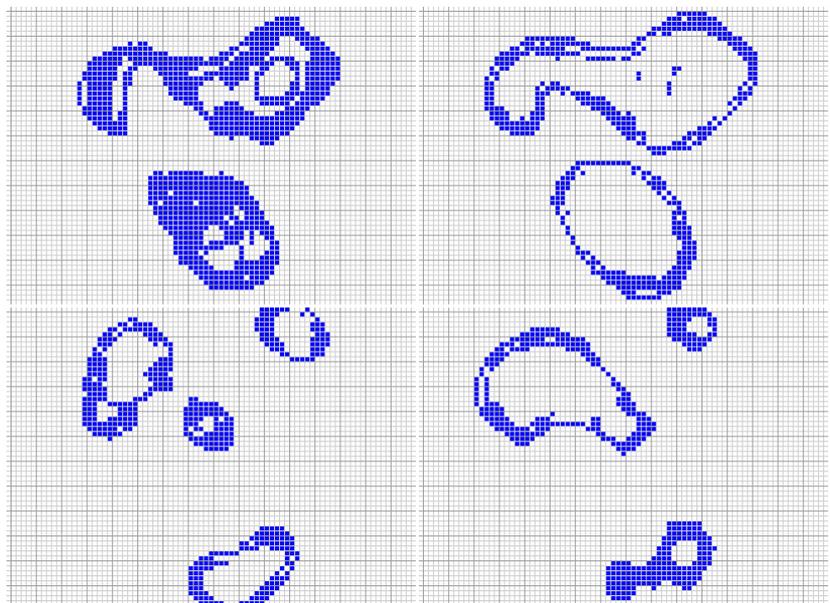

This category IIIa contains also all patterns of the normal "Game of Life": Blinkers, Gliders, Spaceships, etc. In the normal Game of Life from John Conway there are for instance spaceships in all different sizes, lightweight, medium or middleweight and heavyweight, besides the usual gliders, the most basic "spaceships". A spaceship moves with light-speed, i.e. one pixel or unit per time step.

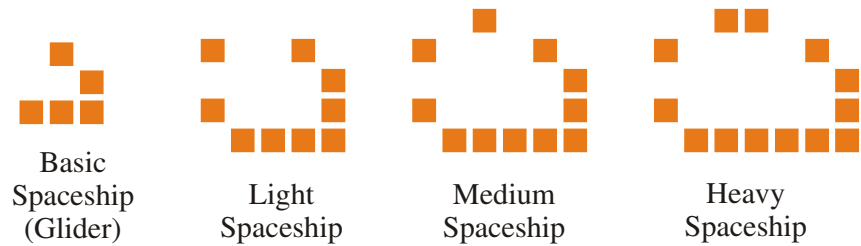

| Basic Spaceship (Glider) | Light Spaceship | Medium Spaceship | Heavy Spaceship |

All these spaceships look similar, and they propagate vertical or horizontal direction through the same mechanism, a repeated contraction and expansion due to coupled positive and negative feedback (except the glider which travels in diagonal direction). The mechanism is a sort of repeated bubbling, the form of the spaceship can be seen as the border of an expanding and contracting bubble.

Spaceship Propagation in the Game of Life

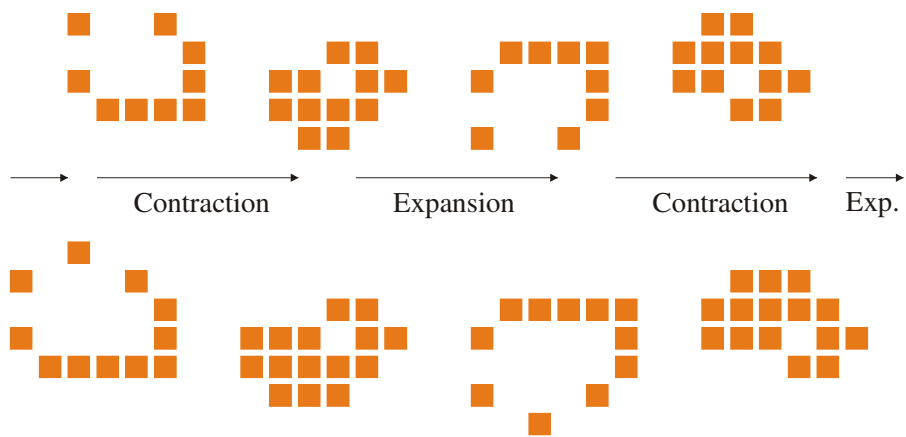

Contraction → Expansion → Contraction → Exp.

### 5.3 b) Tunneling – Adaptive Emergence with multiple feedback

There are many reasons for the slow or sudden appearance of complexity. If something emerges very suddenly or fast, it has for instance often been blocked before by an obstacle or barrier, e.g. a jam, a dogma, or barrier or a system border. A large ecosystem consists of thousands of species and their corresponding ecological niches and habitats, many of them interacting with each other. It usually consists of many different plants, animals, and various micro-organisms like bacteria that are linked by a very complex network. If we consider a global ecosystem like the earth as a whole, we can notice numerous types of feedbacks and constraints which make up together a very complex system. As the history of our Earth shows, the evolution of such a system is certainly not a linear, smooth and continuous process. It is marked by abrupt, unsteady changes and jumps in complexity.

The different levels of complexity and organization in life forms are associated with evolutionary transitions [SmithSzathmáry97]. Evolutionary transitions characterize the crossing of large fitness gaps and fitness barriers. Abrupt, unsteady changes and jumps in complexity are the consequence of unsteady fitness landscapes and barriers. Evolution waits until major events like massive catastrophes break and reduce these fitness barriers or agents are able to tunnel through them. Catastrophes act as catalysts, if they accelerate the transition to higher forms of complexity through a sudden, dramatic increase of challenges in the environment. The tunnelling through fitness barriers is possible through the borrowing of complexity, similar to the borrowing of energy during a tunnelling process in Quantum Theory, see chapter 5.2 and 5.3 in [Fromm04].

This form of emergence in adaptive and evolutionary systems is directly related to (mass) extinctions and dramatic or catastrophic events in the environment. Catastrophes in natural systems can be comets or asteroid impacts, volcanoes, earthquakes, ice ages, droughts or floods. If there are catastrophic events or fluctuations which are unpredictable and neither too common nor too rare, then these catastrophes can enhance evolution and accelerate adaptation. Therefore this form of emergence can be named *adaptive emergence.* It is an example of **type IIIb** emergence, and appears in a complex adaptive system (CAS) with multiple feedbacks and many constraint generating processes. It is associated with the appearance of completely new roles and the dramatic change of already existing "ecological" niches.

**Type IIIb** emergence is also responsible for sudden scientific and mental revolutions. Before "mental revolutions" there is usually a mental barrier for new actions or insights – an incongruity in meaning or an unconscious censor which inhibits actions. Every new situation for a human or an agent is a cognitive puzzle or problem. If it can not be solved at all by thinking and reasoning, it can turn out potentially into a catastrophe. Thus every mental barrier is like a small cognitive catastrophe, which results in a new insight and an avalanche of neural activity (or laughter) if the obstacle is suddenly overcome. The same argument applies to scientific revolutions proposed by Thomas Kuhn: a barrier (often caused by the constraints of the old theory) prevents the discovery of a new theory, until the obstacle is overcome with a new paradigm which results in an avalanche of publications and scientific activities, see chapter 6.2 in [Fromm04].

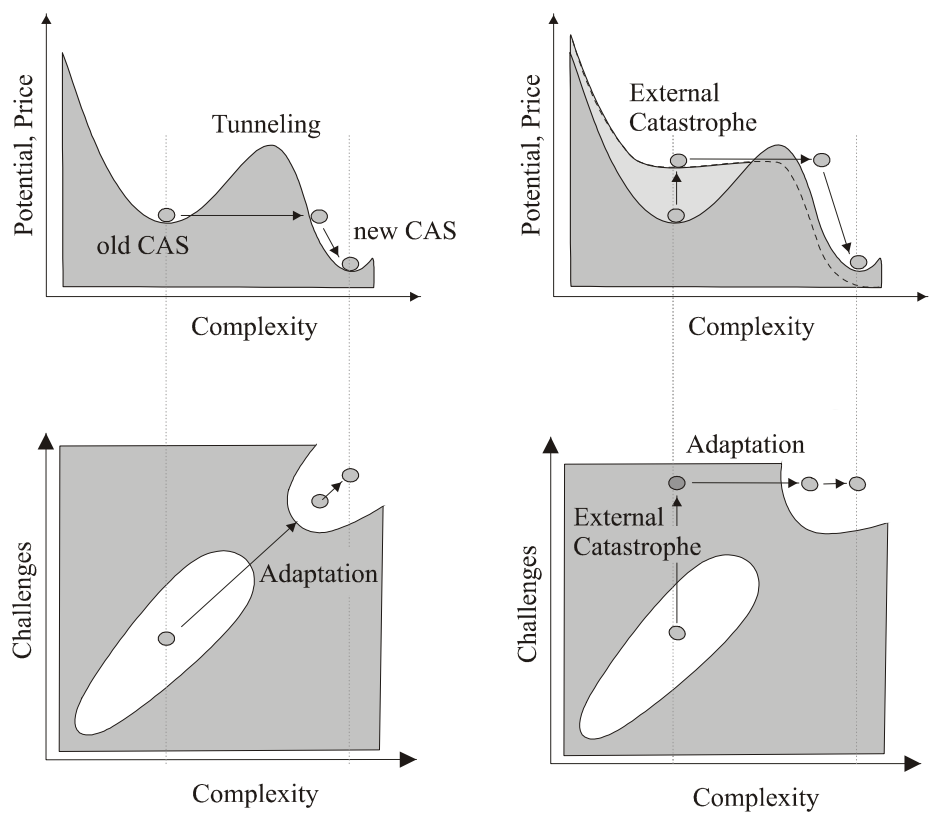

## 5.4 Type IV. Strong Emergence and Supervenience

*Strong Emergence* can be defined as the appearance of emergent structures on higher levels of organization or complexity which possess truly new properties that cannot be reduced, even in principle, to the cumulative effect of the properties and laws of the basic parts and elementary components. **Life** is a strong emergent property of genes, genetic code and nucleic/amino acids, and **Culture** in general is a strong emergent property of memes, language and writing systems. Contrary to the theories of some philosophers, this definition of strong emergence does not have to violate any laws of physics. The term strong emergence is sometimes used to describe magic, unscientific or supernatural processes. This is apparently a wrong concept which must be modified.

The processes which have been described by strong emergence are not magical, unscientific or even anti-scientific. There are no magic or supernatural powers involved, only very complex phenomena on multiple scales. Like other forms of emergence it may look magical, if you do not understand the inner processes. If you have never heard of DNA and genes, then life looks in fact magical. Yet there is a spark of truth in the idea that life can not be explained solely by physical processes ("Vitalism"). It is correct that the physical laws can not describe biological forms. The laws of particles physics are irrelevant to macroscopic phenomena. Microscopic rules are irrelevant compared to the effects of collective organization on macroscopic scales. This is the paradox of emergence, which becomes most clearly visible in the case of strong emergence. The macroscopic structures and patterns depend on the microscopic particles, and yet they are independent from them. This weakest form of causal dependence is also circumscribed by the name supervenience.

The macroscopic level is independent from the microscopic level, because there is a mesoscopic or intermediate level that protects and isolates the one from the other. Therefore in strong emergence the macroscopic level is irrelevant to the microscopic level and vice versa. It is like Anderson said: "the ability to reduce everything to simple fundamental laws does not imply the ability to start from those laws and reconstruct the universe. In fact, the more the elementary particles physicists tell us about the nature of fundamental laws, the less relevance they seem to have to the very real problems of the rest of science, much less to those of society" [Anderson72]. The macroscopic level remains invariant if the micoscopic level is replaced by something else, as long as the mesoscopic or intermediate levels remains the same. Laughlin calls this the "Barrier of Relevance" in his book in chapter 12 [Laughlin05].

Strong emergence in this sense is the *crossing of the barrier of relevance*. It is often related to very large jumps in complexity and major evolutionary transitions, which can be characterized by the appearance of new replicators (genes, memes, ...) and completely new forms of evolution (biological, cultural, ...). The enormous number $10^{120}$ - according to Paul Davies [Davies04] the *Landauer-Wheeler-Lloyd limit* - sets a limit to any constraint of over-determinism that "bottom level" laws of physics might exercise over "higher level", emergent laws, simply due to the problem of combinatorial explosion. The huge number of possible combination makes any deterministic algorithm, rule or law impossible. $10^{120}$ is an astronomical number, which corresponds (very) roughly to the number of bits of information that have been processed by all the matter in the universe (if the universe were turned into a computer, it could process about $10^{120}$ bits of information in the age of the universe, which is roughly 14 billion years [Lloyd02]). Even if a deterministic rule or law is very short and compressed, it makes no sense if the calculation takes several billion years. This does not mean that determinism is completely impossible, it only constrains the determinism of local, low-level laws over global, high-level behavior with strong emergent properties of the system. Examples where the *Landauer-Wheeler-Lloyd limit* is reached definitely are the transitions to the two basic forms of evolutionary systems (genetic and memetic evolution):

(1) **The emergence of life,** biological evolution, genes, genetic replicators, genetic code
Real proteins contain typically between n=60 and n=100 amino acids. The existing 20 amino acids can be arranged in $20^n$ different sequences, which comes close to the Landauer-Wheeler-Lloyd limit of $10^{120}$. Since amino acids in proteins are specified by three letter words or codons (CCG, UGA, CAA, AUG,..) of nucleotide base pairs, the same
argument applies to base pairs in genes.

Some say that this is an argument for "intelligent design" and against evolution. Yet evolution *is* able to explore these vast array of possibilities and combinations over a long period of time. Due to constant variation through mutation and recombination it checks and abandons a huge number of attempts that do not work. The point here is that it is not possible to incorporate the behavior of biological organisms (which are based on genes and proteins) somehow into the laws of atomic physics.

(2) **The emergence of culture,** cultural evolution, memes, memetic replicators, language
Normal alphabetic languages with 26 characters also specify an inconceivable, tremendous large number of different combinations. Languages like English have 26 characters, roughly 7 characters in a word, and about 7 words in a sentence, hence about $26^{49}$
different sequences for a sentence. Not all combinations are valid, though, and there only about 100,000 English words. If we consider longer meaningful texts with more than 25 words, we come again close to the Landauer-Wheeler-Lloyd limit of $10^{120}$. Again, this huge number of combinations prevents any direct control of low-level biological laws over social and cultural phenomena. Neither our languages nor our writing systems are encoded in our genes (we know that genes control emotions, and emotions control high-level behavior, but this is more an indirect control mechanism).

These two forms are the most important kinds of "emergence". The first, the origin of life and the genetic code, triggered genetic and biological evolution. The second, the origin of culture and language, triggered memetic and cultural evolution. The two elementary units of live and culture are the cell and the society, respectively. The first is based on genetic code and DNA, the second on ordinary language and writing systems. Strong emergence can be seen as a transition from one evolutionary system (or complex adaptive system) to another and is therefore a kind of breakthrough or *gateway event*, as Murray Gell-Mann calls it [Gell-Mann94]. A gateway event opens up whole new realms of possibility. Since strong emergence is related to the appearance of a new code, it has the tendency to spawn many different forms of weak emergence. Systems (not individual agents) which are the result of strong emergence are very complex. The role of emergence in both fields, living cells and changing societies, is still an active field of research, see for example [Sawyer05].

Type IV means up to now the appearance of new evolutionary systems where structures are specified by a new language in form of a genetic/memetic code or other alphabet. But in principle it can be any system with a set of building blocks and multi-level emergence. An example is the simulation of a Turing Machine in the Game of Life. Although it may be considered as useless to build a Turing Machine in a CA[6], it can serve as a toy-model for strong emergence, because here the emergent phenomena implement functionality that is logically separable from the underlying phenomena that gave rise to it. The Turing Machine functionality may be implemented on a platform defined by Game of Life rules, but in any other platform as well, and the Computability Theory is not a subset of Game of Life theory – if there is such a thing.

---
[6] It would take nearly an infinite amount of time and resources to do something useful. It is like traveling around the world with a tricycle: awfully slow and not very meaningful.

In philosophy this is described by the term supervenience, a step before "transcendence". *The highest/strongest form of emergence is related to supervenience, the lowest/weakest form of causal dependence*. In other words the maximal form of emergence is associated with the minimal form of causal connection. Supervenient means originally "not part of the real or essential nature of a thing". A "higher-level system X is supervenient on a low-level system Y", if the following holds for all objects a and b: (1) a and b cannot differ in their X-properties without also differing in Y-properties (2) if a and b have identical Y-properties, then they also have identical X-properties. It is often used to explain the brain-mind or body-soul duality, where X are mental properties and Y are physical properties.

Therefore strong emergence describes systems which are supervenient on the lower ones, but logically separable from the underlying phenomena that gave rise to it. "Multiscale" systems arising through strong emergence depend on lower systems, because they are implemented at the lowest level in their language, but at the same time they are independent from them, since they obey their own mesoscopic or macroscopic language and are insensitive to microscopic details. The microscopic details of the lower system are completely irrelevant to macroscopic phenomena. The higher system could be implemented in other systems as well.

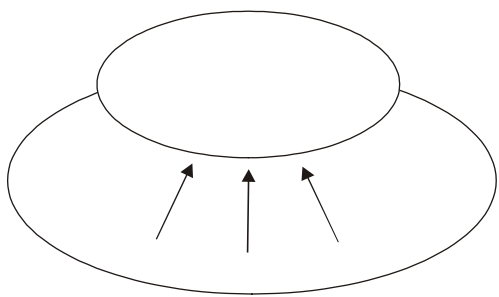

TYPE I
Simple Emergence

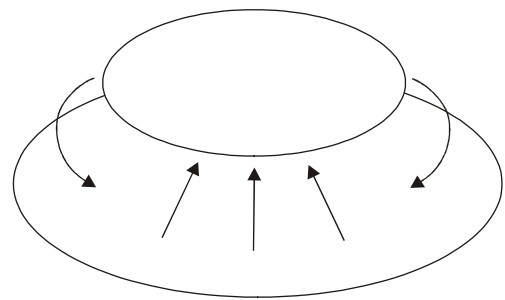

TYPE II
Weak Emergence

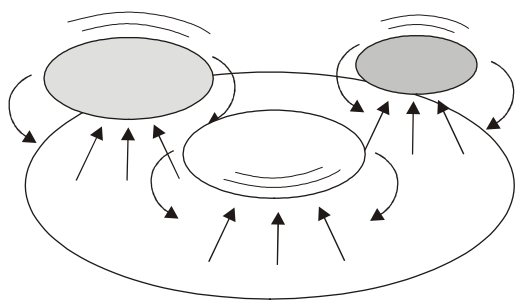

TYPE III
Multiple Emergence

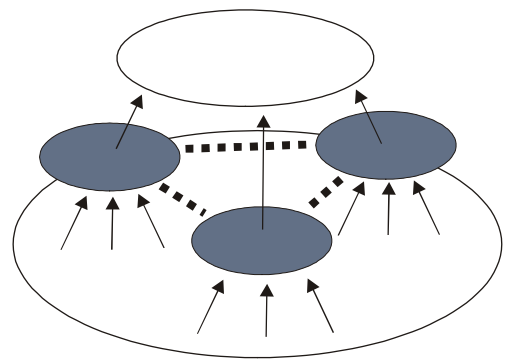

TYPE IV
Strong Emergence

# 6. Conclusions

This paper offers a comprehensive classification for the different types of emergence, which are divided into four basic classes: **Type I** describes simple emergence without top-down feedback and self-organization, and includes esp. intentional emergence in complicated machines. **Type II** contains the classic phenomena of weak emergence including top-down feedback and self-organization. It is further distinguished between stable and instable forms in this class. **Type III** covers all forms of emergence through multiple feedback and adaptation in more complex adaptive systems due to evolution, and finally **Type IV** characterizes all forms of strong emergence in evolution. The term strong emergence is liberated from any magical or unscientific meaning.

The proposed classes stay the same if you select predictability instead of feedback as the main characteristic class feature: it does not matter whether you consider degrees of predictability, types of feedbacks, forms of causality, or the different kinds of constrained generating processes or roles. The classification can also be specified in terms of predictability: intentional emergence of **type I** is predictable and corresponds to fixed roles, weak emergence of **type II** is predictable in principle (though not in every detail) and corresponds to flexible roles, multiple emergence of **type III** is often chaotic or not predictable at all, and is associated with the appearance of new roles and the disappearance of old ones, whereas strong emergence of **type IV** is not predictable in principle, because it opens up a whole new world of new roles. Emergence is a creative, contingent and often unpredictable process: the stronger the emergence, the less predictable are the emergent properties, patterns and structures.

It is possible that there are other forms of **Type III** emergence with "multiple feedback" which have not been mentioned so far. Systems with many forms of feedback are often very complex and not easy too understand. Systems with time-delay feedback can show for example chaotic behavior. This is certainly a place to look, if you want to discover new forms of emergence. The taxonomy of emergence proposed in this paper is not a comprehensive theory, but a first step in the right direction to a deeper and better understanding of the various phenomena in complex systems. Just as complexity is hard to define, because something is complex if it is difficult to describe, emergence is hard to capture with a model or a theory, because during an emergence process new entities appear, which are governed by their own laws.

## Acknowledgements

This paper would not have been possible without the helpful suggestions, positive incitements and thought-provoking ideas from Prof. Russ Abbott (California State University, USA). Many thanks to my colleague Thomas Weise for his thorough proof reading of the paper.

## Image Sources

The photo from the rising earth is a NASA pictures from the Apollo 11 mission, scanned by Kipp Teague (http://www.hq.nasa.gov/office/pao/History/alsj/a11/images11.html). The image of a sprouting plant is a *Canna Bengal Tiger,* used with friendly permission from Triple Oaks Nursery & Herb Garden, PO Box 385 - 2359 Delsea Drive, Franklinville, New Jersey (http://www.tripleoaks.com/). The picture of the Wolf Pack is a Public Domain image from U.S. Fish & Wildlife Service . The other color images are from the Clip Art and Media section of the Microsoft Office Online website.